\IfFormatAtLeastTF{2024-11-01}
  {\DocumentMetadata{tagging=on,
    tagging-setup={math/setup=mathml-SE},
    pdfstandard=ua-2,
    lang=en-GB}}
  {\DocumentMetadata{lang=en-GB}}
\documentclass{article}
\usepackage[T1]{fontenc}
\usepackage[a4paper,top=0.8in, bottom=0.8in, left=0.8in, right=0.8in]{geometry}
\usepackage{hyperref}

\hypersetup{
  pdftitle={{Ten simple rules for non-visual, reproducible and accessible bioinformatics}},
  pdfauthor={Jacqueline G. Kientsch, Stephan Neuhauss, Izaskun Mallona},
  pdfsubject={Accessible and reproducible bioinformatics for blind and low-vision researchers},
  pdfkeywords={accessibility, reproducibility, single-cell RNA-seq, screen readers, PDF/UA},
  pdfdisplaydoctitle=true,
}
\usepackage{comment}
\usepackage{tabularx}

\usepackage{longtable}
\usepackage{array}
\usepackage{ragged2e}
\usepackage{graphicx}
\usepackage{float}
\usepackage{csvsimple}
\usepackage{seqsplit}
\usepackage{xcolor}
\usepackage{fancyvrb}

\usepackage[numbers]{natbib}

\usepackage{orcidlink}

\usepackage{titling}
\posttitle{\par\end{center}\vspace{-0.8em}}
\postauthor{\end{tabular}\par\end{center}\vspace{-2em}}
\pretitle{\begin{center}\singlespacing\LARGE}

\usepackage[none]{hyphenat}
\newcolumntype{P}[1]{>{\RaggedRight\arraybackslash}p{#1}}

\IfFileExists{tagpdf.sty}{\usepackage{tagpdf}}{}
\ifdefined\tagmcbegin
  \newcommand{\arthline}{\noalign{\tagmcbegin{artifact}}\hline\noalign{\tagmcend}}
\else
  \newcommand{\arthline}{\hline}
\fi

\usepackage{setspace}
\usepackage{etoolbox}
\AtBeginEnvironment{tabular}{\singlespacing}
\AtBeginEnvironment{longtable}{\singlespacing}
\AtBeginEnvironment{Verbatim}{\singlespacing}
\AtBeginEnvironment{thebibliography}{\singlespacing}

\newcommand{\fnbrk}{\_\allowbreak}

\providecommand{\tagpdfparaOff}{}
\providecommand{\tagpdfparaOn}{}

\newcommand{\blockgraphic}[2][]{%
  \tagpdfparaOff
  \includegraphics[#1]{#2}\par
  \nointerlineskip
  \raisebox{0pt}[0pt][0pt]{\includegraphics[width=2pt,artifact]{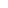}}\par
  \tagpdfparaOn}

\newcommand{\artifactgraphic}[3][]{%
  \tagpdfparaOff
  \includegraphics[#1,artifact]{#3}\par
  \nointerlineskip
  \raisebox{0pt}[0pt][0pt]{\includegraphics[width=2pt,alt={#2}]{figures/white_pixel.png}}\par
  \tagpdfparaOn}

\newcommand{\figonealt}{A five-level pyramid of the ten rules, read from the
  base upward. Each level comes first, then its barriers, then its
  facilitators, and every box names the level it belongs to. Level 1,
  accessible computing environment, holds rule R3. Level 2, readable data and
  objects, holds R5 and R6. Level 3, analysis you can justify, holds R1, R2
  and R4. Level 4, documents and collaboration, holds R7 and R8. Level 5,
  shared infrastructure, holds R9 and R10. Above the apex is the overarching
  goal: reproducible, FAIR and accessible research, carried out
  independently. This figure is browseable: its labels are real text, so you
  can read through them level by level rather than relying on this description
  alone.}

\newcounter{suptab}
\renewcommand{\thesuptab}{Table~S\arabic{suptab}}
\newcommand{\SupTable}[1]{\refstepcounter{suptab}\label{#1}\noindent\textbf{\thesuptab.}}
\newcounter{supfig}
\renewcommand{\thesupfig}{Figure~S\arabic{supfig}}
\newcommand{\SupFig}[1]{\refstepcounter{supfig}\label{#1}\noindent\textbf{\thesupfig.}}

\title{Ten simple rules for non-visual, reproducible \\ and accessible bioinformatics}

\newcommand{\affil}[1]{#1\par\vspace{1pt}}
\author{
  Jacqueline G. Kientsch$^{1,2,*}$~\orcidlink{0009-0008-4253-2140},
  Stephan C.F. Neuhauss$^{1}$~\orcidlink{0000-0002-9615-480X},
  Izaskun Mallona$^{1,3,4,*}$~\orcidlink{0000-0002-2853-7526} \\[4pt]
  \parbox{0.86\textwidth}{\centering\small
    \affil{$^1$ Department of Molecular Life Sciences, University of Zurich, Zurich, Switzerland}
    \affil{$^2$ Life Science Zurich Graduate School, Ph.D. Program in Molecular Life Sciences, Zurich, Switzerland}
    \affil{$^3$ SIB Swiss Institute of Bioinformatics, Zurich, Switzerland}
    \affil{$^4$ Current address: Department of Biosystems Science and Engineering, ETH Zurich, Basel, Switzerland}
    \vspace{3pt}
    \affil{$^*$ Correspondence: \texttt{jacqueline.kientsch@uzh.ch} or \texttt{jacqueline.kientsch@gmail.com}}
    \affil{\phantom{$^*$} \texttt{izaskun.mallona@bsse.ethz.ch} or \texttt{izaskun.mallona@gmail.com}}
    \vspace{2pt}
    \affil{\footnotesize ORCID, in author order: \texttt{0009-0008-4253-2140},
      \texttt{0000-0002-9615-480X}, \texttt{0000-0002-2853-7526}}}
}
\date{}

\begin{document}

\maketitle

\section*{Abstract}
Bioinformatics workflows rely heavily on visual representations. Quality-control plots, cell embeddings, heatmaps, genome-browser tracks, and interactive dashboards are not merely illustrations, but instruments for making analytical decisions. For blind and low-vision researchers who use screen readers, braille displays, or audio-based interfaces, these create a barrier: the evidence used to justify an analysis is often encoded in visual form, while the underlying decision remains undocumented.

We argue that non-visual accessibility and computational reproducibility are closely aligned, as they both require analyses to be transparent and to record why decisions were made. We present ten simple rules for non-visual bioinformatics, covering plots as decision records, cautious use of AI-generated figure descriptions, accessible computing environments, text-first literate programming, structured data and metadata, compact object summaries, accessible publication formats, collaboration practices, shared community infrastructure, and accessibility as part of FAIR research. The intended audience is computational biologists and developers.

Using single-cell RNA-seq as a running example, we show that the accessible equivalent of a plot is a structured decision record. That is, a plot companion that goes beyond storing the underlying data by also stating the purpose of the analysis and the resulting quantitative evidence and uncertainty. We argue that treating accessibility in this way makes bioinformatics more inclusive and also more transparent and auditable.

\section*{Introduction}

Extracting biological insight from computational biology analysis requires inspecting data and analysis results. In single-cell and bulk RNA-seq, analysts examine dimensionality reduction plots to assess clustering and integration or heatmaps to annotate cell identities. Hence, much weight is placed on reading graphics. For blind and low-vision researchers, this creates a barrier, as a rasterised plot is opaque to a screen reader. The larger problem is that the analytical judgement supported by such plots is often not recorded elsewhere: while figure captions may describe what a finished figure shows, they rarely explain which threshold was chosen, which alternatives were considered, which data support the decision, or the associated uncertainty.

This problem is structural rather than anecdotal~\cite{smits2025biomedicalresources}. A recent large-scale evaluation of life-science resources found severe accessibility issues in 74.8\% of data portals and 69.1\% of journal websites. In the same evaluation, manual data discovery tasks carried out with a screen reader reached only a 53.3\% success rate ~\cite{lyi2025lifesciences}. Recommendations for making web-based biomedical data resources accessible exist \cite{smits2025biomedicalresources}. Our focus is complementary, as we address the day-to-day practice of computational analysis itself: inspecting data, justifying thresholds, translating plots and other diagnostic tools into decisions, and exposing the whole process to researchers who do not work visually.

These rules grew out of lived experience, where accessibility barriers persist even in the age of Artificial Intelligence (AI). The first author works as a blind bioinformatician and uses screen readers, braille, and audio feedback to navigate code, data, and scientific literature. The remaining authors learned many of these constraints through collaboration by noticing that reproducible analysis in the computational sense could still be inaccessible to others. Some recommendations are supported by published accessibility research, whereas others reflect personal experience.

Reproducible research requires keeping track of how every result was produced, avoiding manual data manipulation, storing the data behind plots, version-control scripts, and automating the management of software environments~\cite{sandve2013reproducible}. FAIR data stewardship similarly emphasises findability, accessibility, interoperability, and reusability for humans and machines~\cite{wilkinson2016fair}. We believe reproducible research and accessibility are synergistic. Non-visual accessibility applies those principles more thoroughly. For instance, if a result depends on a visual judgement, we suggest writing down that judgement. Similarly, if a figure supports a claim, the same data should be provided as a structured table. In this sense, the rules work as a cognitive device for uncovering implicit, plot-based decisions. The same rules help readers with colour vision deficiency: a decision record (\textbf{Rule 1}) does not depend on hue, and figures should not use colour as the only way to show information (\textbf{Rule 7}).

\begin{figure}[tbp]
\centering
\blockgraphic[width=\textwidth, alt=\figonealt]{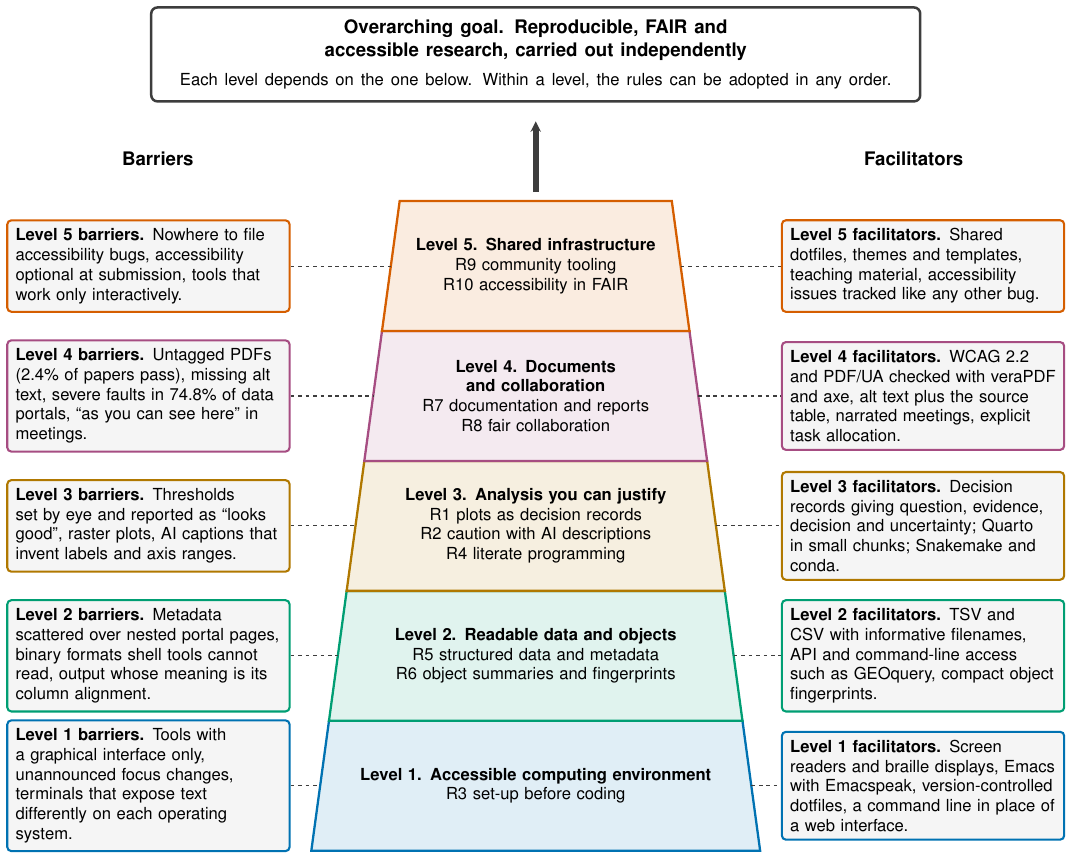}
\caption{The ten rules as a hierarchy of needs. Each level depends on the one below: from the computing environment, which a researcher must be able to start, navigate, and operate with assistive technology before any analysis is possible
(level~1) otherwise never reaches the data (level~2), and so on up to shared infrastructure (level~5). Beside each level are the barriers we have met there and the facilitators that remove them. Rules keep the numbering used in the text, abbreviated as R1 to R10. The rules are numbered in order of argument and arranged in the figure by dependency. The labels are text rather than pixels, and are ordered so that a screen reader reads each level, then its barriers, then its facilitators, ending at the overarching goal.}
\label{fig:overview}
\end{figure}

We use single-cell RNA-seq (scRNA-seq) data analysis as the running example because it is a complex and plot-heavy task. Standard scRNA-seq workflows involve quality control, normalisation, dimensionality reduction, clustering, and annotation~\cite{luecken2019bestpractices}. At nearly every step, researchers inspect plots and make decisions. To show how to incorporate accessible and reproducible practices into the workflow, we provide an accessible analysis of the commonly used 3k PBMC dataset~\cite{tenxpbmc3k} as analysed by Seurat~\cite{hao2024seurat,seuratpbmc3k}, demonstrating decision records along the analysis (S1 File). Figure~\ref{fig:overview} arranges the ten rules as five levels of need, from the computing environment a researcher has to be able to drive at all, up to the shared infrastructure that keeps the practice going and depends on institutional support, together with the barriers and facilitators we have met at each level.

\section*{Rule 1: Treat plots as decision records, not illustrations}

In computational biology, plots are often used to develop an intuition about data structure and underlying biology. Ideally, this intuition should agree with conclusions reached through empirical evaluation. The data supporting that evaluation can be made accessible, which makes the analysis transparent and easier to comprehend.

Supporting data should go beyond using long captions in figures, and should be closer to decision records, keeping track of the purpose of the diagnostic, identifying the data being inspected, reporting the numerical evidence, flagging uncertainty, recording the decision taken, and preserving the data table from which the plot was generated. This makes the analysis accessible to blind and low-vision researchers, but it also makes it explicit for any reader, including collaborators or the original analyst returning months later.

A decision record structures information at several levels. First, it states the analysis aim: for example, whether a sample has acceptable quality, whether integration removed batch effects without erasing biological signal, or whether a cluster label is supported by marker expression. Second, it reports the evidence in text and tables: quantiles, thresholds, effect sizes, enrichment statistics, etc. Third, it records the decision and its rationale. Fourth, it preserves ambiguity and uncertainty, if present. For instance, if two cluster labels remain plausible, or if a threshold is borderline, it keeps track of that uncertainty. We propose a full set of properties for decision records in~\ref{stab:adr}.

The purpose of such records is to enrich statements such as ``The UMAP looked good after integration.'' with ``After integration, donor-associated separation decreased while known cell-type marker structure was preserved. No major cluster was dominated by a single donor. The selected integration was retained for downstream clustering, but small donor-specific differences in cluster 7 should be rechecked during annotation.'' The latter lets a reader retrace the results and the reasoning behind them.

\begin{figure}[htbp]
\centering
\artifactgraphic[width=0.6\textwidth]{Three violin plots show detected genes, total UMI counts, and mitochondrial percentage before filtering.}{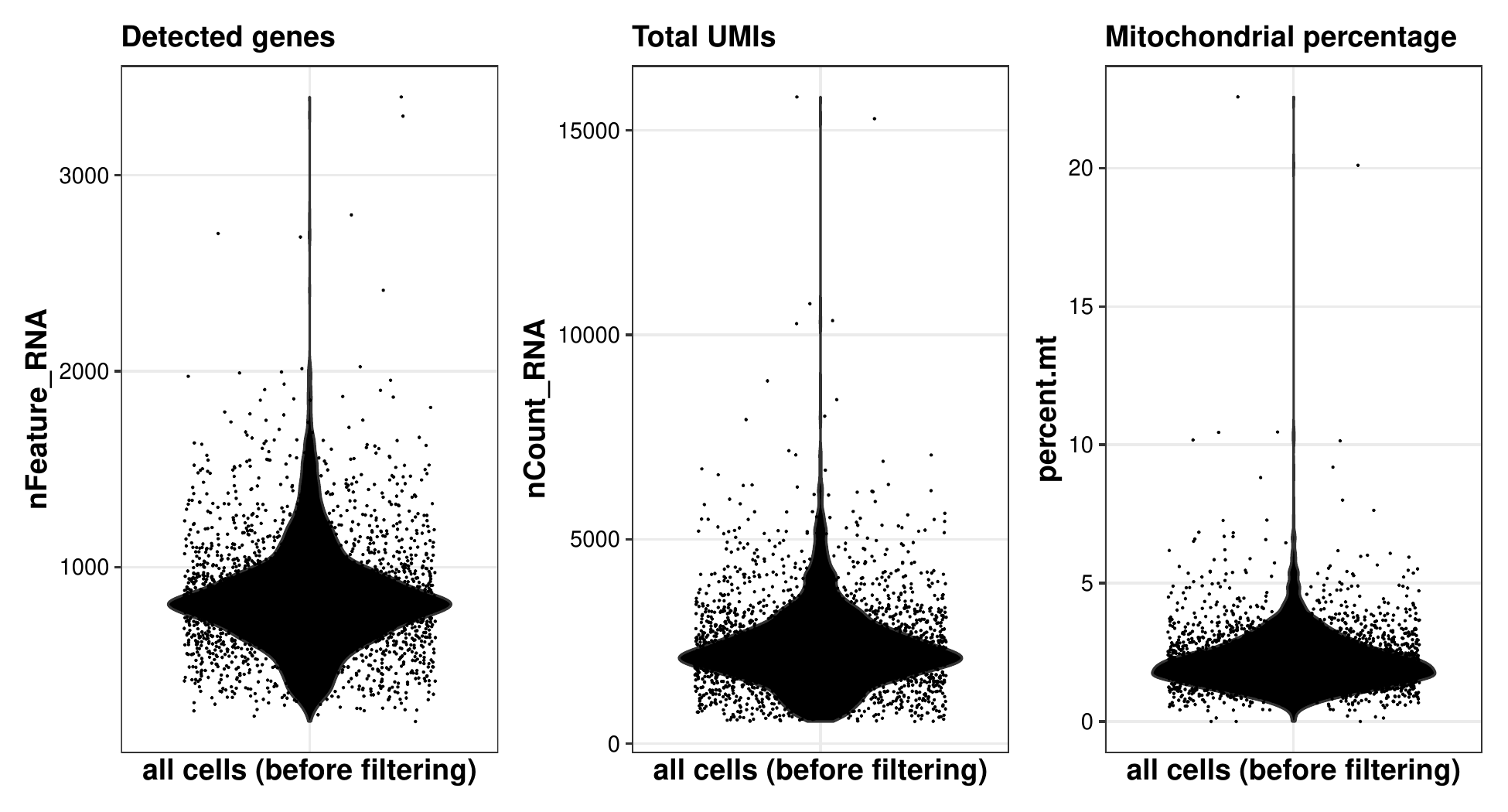}
\caption{Quality-control violin plot: the distributions a sighted analyst reads to judge cell quality and set thresholds. Its non-visual companion is Table~\ref{tab:qc_diagnostic}.}
\label{fig:qc_diagnostic}
\end{figure}

We exemplify a decision record with one of the first steps of quality control (QC) in scRNA-seq data: filtering in cells suitable for downstream analysis. The task is to inspect the distribution of the gene number, the number of unique molecular identifiers (UMIs), and the mitochondrial percentage, and empirically selecting the thresholds to flag low-quality cells. Distributions are traditionally plotted as violin plots (Figure~\ref{fig:qc_diagnostic}) and thresholds set by visual inspection. We propose to describe quantiles instead (Table~\ref{tab:qc_diagnostic}) and to record the decision in the caption of that table, with five of the required fields of~\ref{stab:adr}.

\begin{table}[htbp]
\centering
\caption[Non-visual companion of the quality-control violin plot and decision record for cell filtering.]{Non-visual companion of the quality-control violin plot (Figure~\ref{fig:qc_diagnostic}). The same three distributions as exact quantiles, with the number of cells each cutoff removes, which the plot only approximates. Values from S1 File (\texttt{02\_qc\_distribution\_before\_filtering.tsv} and \texttt{02\_qc\_failure\_summary.tsv}).
\par Decision record, with five of the required fields of~\ref{stab:adr}; the example column of~\ref{stab:adr} gives the full record.
\par\textit{Question:} Which cells should be kept for downstream analysis?
\par\textit{Decision:} Keep cells with more than 200 and fewer than 2{,}500 detected genes and less than 5\% mitochondrial counts.
\par\textit{Rationale:} The upper cutoffs remove the high tails of detected genes (5 cells) and mitochondrial percentage (57 cells), and they match the official tutorial~\cite{seuratpbmc3k}.
\par\textit{Uncertainty:} Fixed cutoffs are heuristics, and the lower gene cutoff removes no cells. Outlier detection at three median absolute deviations flags 173 cells, including 57 of the 62 removed cells and 10 of the 13 platelets (\ref{stab:adr}).
\par\textit{Downstream effect:} 2{,}638 cells are normalised and used in all later steps.}
\label{tab:qc_diagnostic}
\small
\begin{tabular}{l r r r r r}
\arthline
\textbf{Metric} & \textbf{Min} & \textbf{Q25} & \textbf{Median} & \textbf{Q75} & \textbf{Max} \\
\arthline
Genes & 212 & 690 & 816 & 952 & 3{,}400 \\
UMIs & 546 & 1{,}756 & 2{,}196 & 2{,}762 & 15{,}818 \\
Mito \% & 0.0 & 1.5 & 2.0 & 2.6 & 22.6 \\
\arthline
\end{tabular}

\medskip
{\footnotesize Cells removed by each cutoff: 5 above 2{,}500 genes (0.19\%), 57 above 5\% mitochondrial (2.1\%), 62 failing at least one (2.3\%).}
\end{table}

Table~\ref{tab:checkpoints} lists plot types common in bioinformatics, a general non-visual companion for each, their typical uses, and the matching scRNA-seq checkpoints.

\begin{table}[htbp]
\centering
\caption{Common plot types and their non-visual companions. The general companion applies to any task; the last column gives the scRNA-seq checkpoint and its task-specific companion.}
\label{tab:checkpoints}
\small\singlespacing
\begin{tabularx}{\textwidth}{P{0.16\textwidth} P{0.29\textwidth} P{0.18\textwidth} >{\RaggedRight\arraybackslash}X}
\arthline
\textbf{Plot type} & \textbf{General non-visual companion} & \textbf{Typical uses} & \textbf{scRNA-seq checkpoint and companion} \\
\arthline

Scatter plot & Number of points, range and quantiles identities of labelled outliers (all scatter plots); correlation (if two continuous variables); description of other data layers (e.g. point size, colour encoding, point symbol)  & Agreement between replicates and readouts; volcano plots; bubble charts & Quality control: correlation of UMIs with detected genes, cell embeddings, feature plots, dot plot \\

Distribution plot (violin, box, histogram, density) & Number of observations per group, minimum, quartiles, median, maximum, median absolute deviation, number and fraction of observations beyond each threshold, number of modes & Quality control of samples, cells or reads; distributions of scores & Quality control: per-sample number of genes and UMIs, retained and removed cells (Table~\ref{tab:qc_diagnostic}), doublet scoring. \\

Heatmap & Rows and columns in plotted order, value scale, top features per group with effect sizes, reasoning behind ordering (e.g. clustering or not) & Expression across samples or clusters, correlation matrices, multimodal data representation & Annotation: top markers per cluster. \\

Ordered curve & Rank and value at each inflection point, value range, trends, curve shape & Cell calling, sequencing saturation, melting curves, genome browser tracks, time series & Cell calling: knee plot for cell barcodes, scree plot for dimensionality reduction. \\

\arthline
\end{tabularx}
\end{table}

\section*{Rule 2: Take AI descriptions with a grain of salt}

As of November 2026, multimodal large language models are now producing textual descriptions of figures, with integrations for screen readers and document viewers. The descriptions are sometimes excellent, and they have already improved the usability of legacy raster figures that lack alternative text (alt text). They address part of the plotting limitations described in Rule~1, but they are not yet reliable enough to stand alone as scientific captions: in a large annotated set of chart captions, even advanced models often produced factual errors~\cite{huang2024chartcaptions}. In our experience, common failure modes include hallucinated labels, mis-read axis ranges, and confident descriptions of trends not backed up by the data.

Computer-use agents, which operate graphical interfaces through clicks and keystrokes on a user's behalf, could perform some interface tasks for the user that are hard to do with a screen reader (\textbf{Rule 8}). Current agents still fail on many of these tasks. In one evaluation, an agent completed 78.3\% of everyday tasks by default, but only 41.7\% when restricted to keyboard input~\cite{gubbimohanbabu2026a11ycua}. In a three-week diary study with eight blind users, the best model completed 52.5\% of commands~\cite{kodandaram2026cua}.

In the absence of underlying code and decision records (\textbf{Rule~1}), we find AI descriptions helpful only as a first-pass aid, as they cannot be fully verified. We recommend cross-checking these descriptions by comparing outputs from different models and against colleagues' assessments, as recommended in general guidance on AI use~\cite{auslander2026ai}. This provides a workaround without fully solving the problem. In the worked example, the alt text of each figure is written from the plot's data and code (\ref{stab:sup_equivalence}). \textbf{Rules 3--10} describe actionable and reproducible strategies to overcome these limitations.

\section*{Rule 3: Set up an accessible environment first}

Before carrying out any analysis, the operating system (OS) should already manage a screen reader, a braille display, or magnification predictably. Screen readers vary in features and OS compatibility~\cite{gaggi2019accessibility}; we elaborate on the strengths and limitations of five of them, which vary in availability and cost (\ref{stab:screenreaders}).

On macOS, \texttt{VoiceOver} is mature and tightly integrated. On Linux, \texttt{Orca} is fully open and scriptable but integrates fewer applications. On Windows, \texttt{NVDA} is free, open source, widely used, and especially common in many research and educational settings, while \texttt{JAWS} remains heavily used and feature-rich but commercial \cite{webaim2024srsurvey}. In our daily bioinformatics work \texttt{NVDA} and \texttt{JAWS} have often provided the most practical balance of capability, scriptability, and availability. \texttt{Narrator} is improving but still trails both on developer tooling. The three do not respond the same way to the same content. Braille displays operate through the screen reader, so they inherit whatever the OS layer can or cannot expose.  

In terms of terminals and terminal emulators, accessibility depends on the OS and on the specific application. macOS Terminal, GNOME Terminal, Windows Terminal, and Windows Subsystem for Linux do not expose text to assistive technologies in quite the same way. Ideally, these should offer predictable keyboard navigation, clear and stable headings, consistent interfaces, and as few unannounced focus changes as possible.

 The editor choice has a strong impact on day-to-day productivity. Terminal-only tools (\texttt{vim}, \texttt{nano}, and the like) expose text through the same channel as everything else in the shell. Taking R code development as an example, \texttt{Emacs} with \texttt{Emacspeak} \cite{raman1996emacspeak, raman1997aui}, augmented by \texttt{Emacs Speaks Statistics}~\cite{rossini2004ess}, provides a complete set of features, including self-voicing semantics for code, output, and documentation. Visual Studio Code is generally accessible and ships with a screen-reader mode, although side panels, modal dialogs, and unexpected focus changes still cause friction. RStudio has an official screen-reader guide hosted by Posit at \texttt{support.posit.co} \cite{positrstudioaccessibility}; however, several of its panels and dialogs are still unreliable, and the accessibility of Positron, its emerging successor, is not yet clear. Notebooks and R markdown, along with the Python side of this tooling, are discussed in \textbf{Rule 4}.

Once the environment reaches a workable state, we found it useful to keep curated editor settings, key bindings, and SSH profiles under version control as dotfiles, so the same setup can be restored on a new machine instead of being rebuilt by hand. The worked example runs entirely from a shell, and each code chunk can be sent line by line from an editor to a terminal (\ref{stab:sup_equivalence}).

\section*{Rule 4: Lean on literate programming for reproducibility}

Literate programming places explanations and code in the same source file, which after rendering also contains results. For a non-visual reader, a single screen-reader pass through a Python notebook~\cite{kluyver2016jupyter} or R~Markdown~\cite{xie2018rmarkdown} document resembles the visual inspection by a sighted reader.

We find that two practices make this work better. First, keep code chunks small and atomic with one intention (analysis aim) per chunk (code snippet) and a short sentence preceding it. This helps to specify natural analysis boundaries, where a screen reader can stop, and facilitate resuming or re-running chunks on demand. Second, use formats with plain text source, such as \texttt{Quarto} markdown~\cite{allaire2024quarto} or R markdown. Jupyter notebooks' interactivity makes navigation and code execution less intuitive. A study of 100,000 computational notebooks found barriers in the notebook interface itself, in the way data are represented, and in the output that common libraries emit~\cite{potluri2023notably}. Recent versions of JupyterLab and the VS Code Jupyter extension have improved screen-reader support.

To show literate programming in scRNA-seq we provide an accessible and documented \texttt{Quarto} markdown \cite{kientsch2026workflow}.  It implements the recommendations above. The rendered document, in both HTML and PDF formats, including explanations, code, and results, is available as S1 File. Throughout the report, we orchestrate the software environment and execution using \texttt{Snakemake} as a workflow manager~\cite{koster2012snakemake,molder2021snakemake}, archived at Zenodo~\cite{kientsch2026workflow}.

\section*{Rule 5: Structure data and metadata for non-visual exploration}

Structured data and accessible data are not necessarily synonymous. For instance, a web portal can contain well-typed metadata and still scatter information across nested pages, tables, frames, or other components that are hard to navigate by keyboard~\cite{lyi2025lifesciences}. In our experience, command line interface (CLI) tools and application program interfaces (APIs) can reduce the burden encountered on browsing web portals, by making structured data queries easier. For instance, an analysis often starts with retrieving raw data from Gene Expression Omnibus (GEO)~\cite{edgar2002geo}. \texttt{GEOquery}~\cite{davis2007geoquery}, an R package, handles this download through the terminal, bypassing web portal interactions and providing a screen-reader-friendly text object.

After retrieval, data are often stored as files. For tabular data, well-formed tab-separated value (TSV) and comma-separated value (CSV) files are easy to inspect with shell tools and easy to read aloud row by row. Good practices in file naming are also key to browsing directories. We recommend to avoid spaces, stick to lowercase, and put the most distinguishing identifier near the beginning of the name. Binary formats are harder to inspect with shell commands like \texttt{cat}, \texttt{less}, and \texttt{head}, even when (as with Parquet, Arrow, or HDF5) they carry a schema that is rich and machine-readable. In metadata tables, the columns that matter most (sample identifier, condition, batch) sit better on the left. \ref{sfig:tables_listing} shows part of the worked example's output directory of named tables.

Similarly, after a completed analysis, data is often deposited on a FAIR data repository like Zenodo, for which the \texttt{zen4R} R package \cite{zen4R} eases data retrieval and deposit.

\section*{Rule 6: Facilitate data inspection and fingerprinting}

Beyond files themselves, data are inspected during interactive coding sessions and in literate reports. Summaries in R and Python are designed for visual scanning. The default \texttt{print} function produces dense columnar output whose meaning depends on alignment, alignment that is lost during screen-reader-based access. Accessible bioinformatics needs ways to fingerprint data and data objects. We believe the community would benefit by adding alternative \texttt{print} methods with adjustable verbosity. Properties that are particularly useful during object fingerprinting include the in-memory size, the count of missing values, the range and shape of numeric columns, and a one-line characterisation of categorical columns. For very large objects, the same fingerprint is more useful than any sample of rows. In scRNA-seq analysis, we believe object fingerprinting functions should also include cell number, feature number, assay names, layer names, dimensionality reductions, samples, donors, conditions, cluster labels, and basic QC summaries (\ref{stab:objstate}). 

Beyond data objects, Unix \texttt{man} pages exemplify how to structure information in a useful and predictable way: their section headings (NAME, SYNOPSIS, DESCRIPTION) let a screen-reader user jump directly to the part they need. As for predictability, we suggest that argument names be consistent and guessable within and across functions. R offers a cautionary example: the same operation, discarding missing values, is spelled \texttt{na.rm} in \texttt{mean()} but \texttt{useNA} in \texttt{table()}.

\section*{Rule 7: Publish in accessible formats}

                                                        Along with data handling, reports, publications, and documentation are another source of friction. The canonical baseline for accessible digital content is the W3C's Web Content Accessibility Guidelines 2.2~\cite{wcag22}, which covers text alternatives, navigable structure, and predictable interaction. For PDFs specifically, the PDF/UA standard (ISO~14289) defines what a tagged, accessible PDF should provide. That includes a logical reading order, structural tags for headings and tables, and text alternatives for figures. The standard comes in two parts, and the difference matters when choosing a target. Part 1 (PDF/UA-1)~\cite{iso142891} builds on PDF 1.7. Part 2 (PDF/UA-2)~\cite{iso14289} builds on PDF 2.0, so it admits the PDF 2.0 tag set and structure namespaces, applies stricter tag nesting, and disallows features that PDF 2.0 deprecates. It also drops the heading-level requirements of Part 1 and adds requirements for MathML, annotations, and metadata. The Matterhorn Protocol turns the Part 1 requirements into concrete, machine-checkable conditions~\cite{matterhorn2021}; it has no Part 2 counterpart. Validators such as \texttt{veraPDF} cover both parts. Our worked example targets PDF/UA-1, which currently has the wider validator and reader support.

Reports from literate programming practices (\textbf{Rule 4}) tend to work better. We recommend showing code alongside results and including a table of contents as well as hyperlinks to section headings. LaTeX documents using \texttt{tagpdf}~\cite{tagpdf}, and \texttt{Quarto}, \texttt{rmarkdown}, \texttt{bookdown}~\cite{xie2016bookdown}, and \texttt{pagedown}~\cite{pagedown} produce structured and screen-readable documents. Filling in alt text for every figure is, in our experience, the single most useful habit. For readers with colour vision deficiency, figures should use palettes that stay distinguishable under its common forms~\cite{crameri2020colour}, such as Okabe-Ito~\cite{okabeito2008cud} or viridis~\cite{garnier2023viridis}. They should not rely on colour alone, as enforced by the WCAG success criterion 1.4.1 ~\cite{wcag22}.

Scientific PDFs remain a major problem. Wang and colleagues assessed 11,397 scientific PDFs published between 2010 and 2019 and found that only 2.4\% satisfied all of their defined accessibility criteria~\cite{wang2021scia11y}. 

To tackle this problem, we recommend generating both accessible reports and preprints. Our S1 File includes a screen-reader-friendly PDF and HTML report by incorporating PDF tagging, hyperlinks, alt text figure descriptions, screen-reader navigable tables, and embedded code and object fingerprints. As for preprints, the same rules apply, plus avoiding multi-column layouts and line numbers, since they garble the text flow using screen-readers. \texttt{AltGosling} generates easy-to-navigate text descriptions directly from the plot specifications (encoded with \texttt{Gosling}~\cite{lyi2022gosling}), rather than from a rendered image \cite{smits2024altgosling}. More generally, a structured description tied to the underlying data lets screen-reader users explore a figure instead of hearing a single caption~\cite{zong2022richscreenreader}. Each figure should also have the underlying data available as a table (Tables~\ref{tab:qc_diagnostic}, \ref{tab:elbow_diagnostic}, and \ref{tab:umap_diagnostic}). \ref{sfig:accessible_rendering} shows the settings the worked example uses to make its own outputs accessible.

Code repositories deserve the same care. We recommend including a README file and installation instructions.

\section*{Rule 8: Collaborate with accessibility in mind}

Guidelines exist for collaborating effectively~\cite{vicens2007collaboration}. Project completion benefits from clear goals, roles, and timelines, all shared in an accessible text format. Beyond that baseline, we have found that specific practices either ease or obstruct collaboration.

To successfully start a project, data resources have to be shared through channels that work with each collaborator's assistive technology and follow institutional policies on data sovereignty and confidentiality, such as institutional file-sharing services or data repositories. After that, resource allocation matters most. First, blocking factors have to be spelled out, setting boundaries on tasks that are not addressable with current accessibility solutions or workarounds. This also applies to tasks where the effort required to handle a poorly designed graphical user interface (GUI) is disproportionate. For instance, interaction with GUIs with no API could better be assigned to the sighted collaborators. Similarly, sighted collaborators should take responsibility on tasks requiring visual-only evaluation.

Recommendations on how to generate reproducible and accessible results are listed in \textbf{Rules 4, 5, and 6}. As for synchronous communications, we note that plots and any other information shared or projected on screens should be narrated, avoiding ``as you can see here'' during meetings. For staying in touch, keyboard-navigable written exchanges via email and asynchronous messaging (e.g., Slack) are reliable.

Throughout the analysis, to keep track of the outputs exchanged with other team members, we recommend starting the file names with a pointer to the code that generated them, as in the worked example (\ref{sfig:tables_listing}).

Collaborative manuscript drafting benefits from venues where multiple users can edit the document instead of sending around multiple versions of it. A common bottleneck is handling scientific bibliography. In our experience, LaTeX tends to work well (e.g., via Overleaf), keeping references as BibTeX records. 

\section*{Rule 9: Build community and shared infrastructure}

The advice above agrees with the current gold standard for reproducible science: structured and findable data and metadata and good documentation to make analysis traceable. Each rule suggested some improvements to these practices by asking for small changes leading to a big improvement in accessibility. We believe these habits can be adopted more generally, facilitating the dissemination of bioinformatics materials in research but also in teaching. Community efforts exist to provide accessible documents (\textbf{Rule 7}), including validation tools to evaluate the screen-reader compatibility of HTML and PDF outputs. \texttt{veraPDF}~\cite{verapdf} checks a PDF against the machine-checkable conditions of the Matterhorn Protocol~\cite{matterhorn2021}, and \texttt{axe}~\cite{axecore} checks a served HTML page against WCAG~2.2~\cite{wcag22}. Both run without a graphical interface, so they can be coupled to report-generating workflows instead of being run by hand. S1 File reports are generated via a \texttt{Snakemake} workflow where these validations are built in \cite{kientsch2026workflow}.

In terms of potential community-driven and infrastructure developments, we envision shared dotfiles and editor configurations for accessible R, Python, and shell environments (\textbf{Rule 3}), screen-reader-friendly vignettes (\textbf{Rule 4}), example datasets with audited textual summaries (\textbf{Rule 5}), packages and libraries to facilitate non-visual data inspection (\textbf{Rule 6}), and accessibility-aware scientific publication recommendations in journals and preprinting venues (\textbf{Rule 7}). It would also cover teaching material for new blind and low-vision students entering bioinformatics, and a shared place to file and track accessibility issues against bioinformatics software. The worked example ships two small, reusable components of this kind: a stylesheet (\texttt{custom.scss}) that raises link contrast to meet WCAG and wraps long output in the HTML report, and a \texttt{ggplot2}~\cite{wickham2016ggplot2} theme (\texttt{accessible\_theme.R}) that gives every figure large fonts, labelled axes, and a colourblind-safe high-contrast palette for low-vision readers.

Community initiatives can prototype such tools and practices, but maintaining them over years requires institutional commitment. Shared infrastructure for accessible analysis should therefore be part of institutional research support and funded as such.

Similar interest groups exist in computer science, such as the Debian project accessibility community \cite{debianaccessibility}.

\section*{Rule 10: Incorporate accessibility to make FAIR research fair}

The way accessibility is looked at in FAIR data are  grounded on data preservation and retrieval. These rules emphasise a complementary aspect where small incremental changes lead to a large improvement on how usable research materials and findings are for blind and low-vision researchers.

Collaborative research teams should identify accessibility barriers at the start of a project and agree on how data, code, results, and decisions will be exchanged in accessible formats (\textbf{Rule 8}). We note that non-visual accessibility tools are not necessarily tied to reproducibility. Interactive tasks, such as genome browsing for example, are hard to back-track yet can be enabled by other FAIR practices.  VIJB, for example, renders the multilayered output of the JBrowse genome browser~\cite{diesh2023jbrowse} on a braille display, keeping it usable without sight \cite{nashed2026vijb}. In the worked example, the plain-text tables and logs written for non-visual use can also be compared between runs and archived (\ref{stab:sup_equivalence}).

Making an analysis accessible also makes it more explicit and more reproducible.

\section*{A worked example: non-visual exploration of single-cell RNA-seq}

We demonstrate the guidelines above in S1 File, an accessible \texttt{Seurat} analysis of the 10x Genomics PBMC 3k dataset~\cite{tenxpbmc3k,seuratpbmc3k}. It adapts the standard tutorial, but pairs every figure with a non-visual decision record: a table of the plotted values, a short numeric summary of the conclusion, and the underlying data as a downloadable file. 

\begin{figure}[htbp]
\centering
\artifactgraphic[width=0.4\textwidth]{An elbow plot of decreasing principal-component standard deviations that levels off from about the seventh component.}{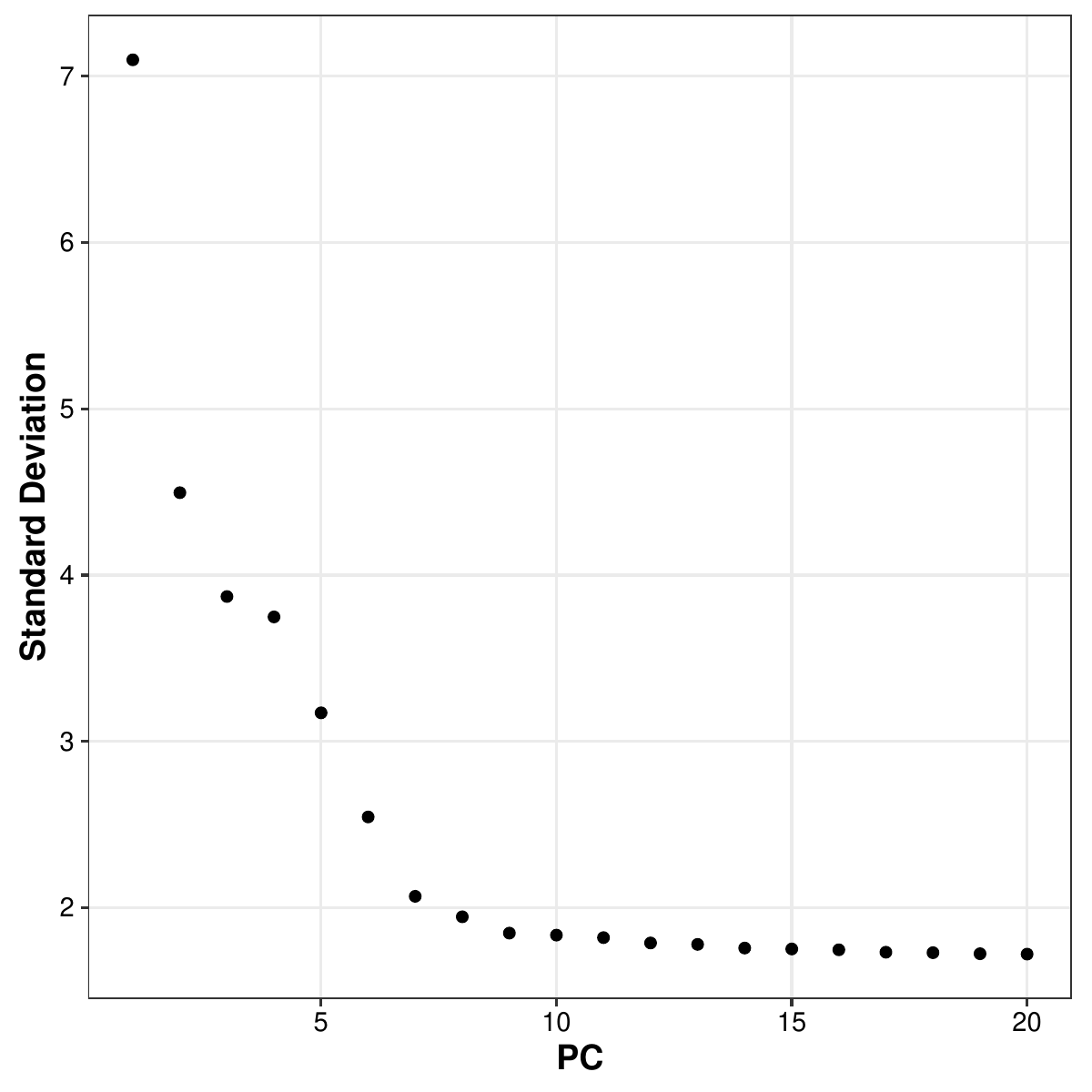}
\caption{PCA elbow plot, read by eye to choose how many principal components to keep. Its non-visual companion is Table~\ref{tab:elbow_diagnostic}.}
\label{fig:elbow_diagnostic}
\end{figure}

\begin{figure}[htbp]
\centering
\artifactgraphic[width=0.7\textwidth]{A UMAP scatter plot with cells coloured and labelled by the nine official PBMC 3k cell types.}{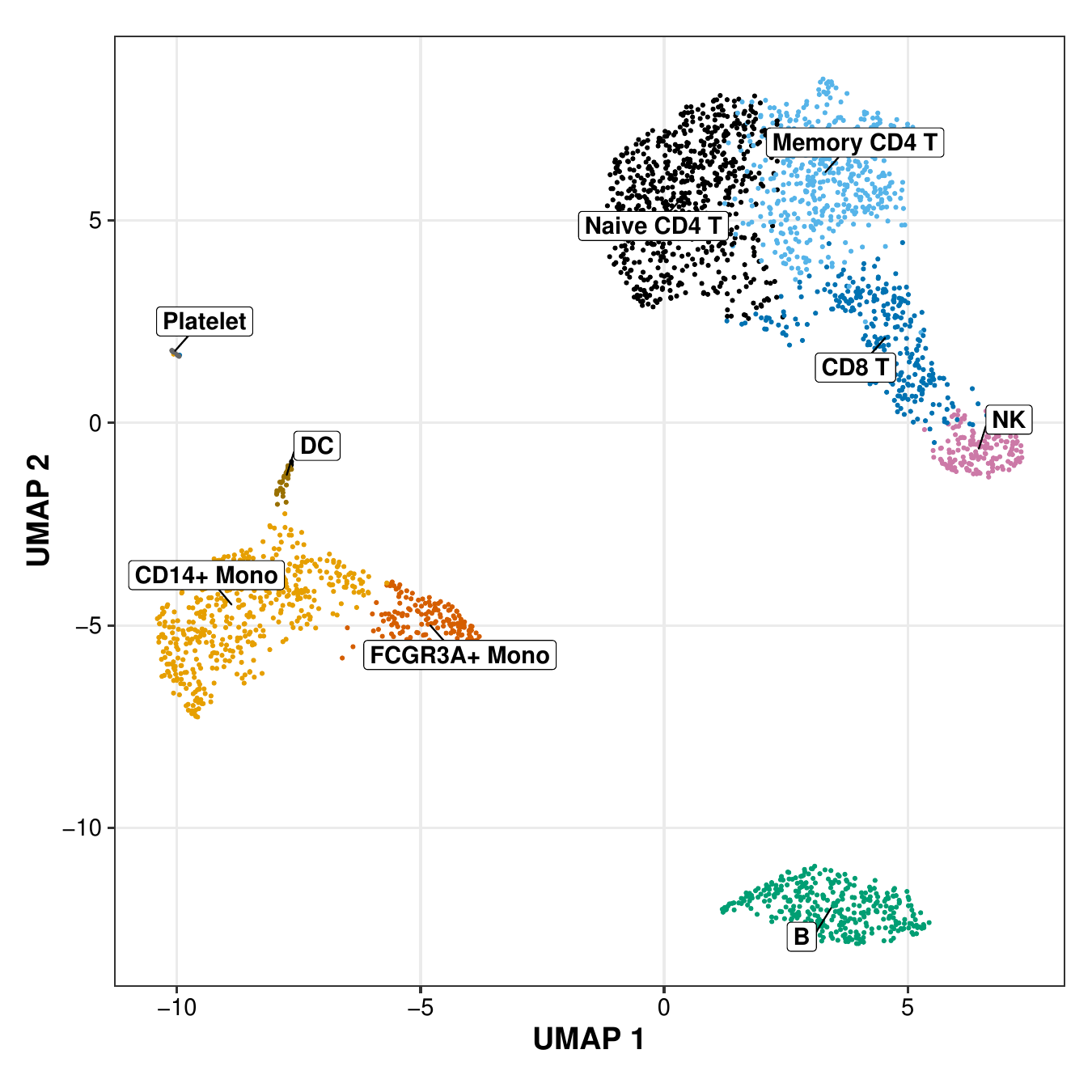}
\caption{UMAP of the PBMC 3k cells laid out by UMAP and coloured by cell type. Its non-visual companion is Table~\ref{tab:umap_diagnostic}.}
\label{fig:umap_diagnostic}
\end{figure}

To orchestrate a reproducible analysis, we provide a \texttt{Snakemake} workflow, including the software environment management via \texttt{conda}~\cite{conda,gruning2018bioconda} (\ref{sfig:reproducible_env}, \ref{stab:params}; \textbf{Rule 4}). The workflow is self-contained, and runs the download and the analysis. It writes a log for every step, so the run can be inspected as text. It marks each finished rule with \texttt{success}, so a reader can confirm a step finished without a plot or a progress bar. It strips the colour codes and non-ASCII characters from the tool output, so the logs read cleanly under a screen reader (\textbf{Rules 3 and 6}). The report is written in \texttt{Quarto} markdown and rendered in two formats, an HTML page and a tagged PDF/UA file, both provided in S1 File. Both keep the headings, tables, and figure descriptions allowing easy keyboard navigation (\textbf{Rule 7}). A custom stylesheet raises link contrast and wraps long output instead of hiding it in a scroll box. Every PDF build is validated with \texttt{veraPDF}, and a separate preview rule checks the HTML with \texttt{axe} (\textbf{Rule 9}). The workflow is available on Zenodo~\cite{kientsch2026workflow}. Figures~\ref{fig:qc_diagnostic}, \ref{fig:elbow_diagnostic}, and \ref{fig:umap_diagnostic} and their matched screen-reader-friendly Tables~\ref{tab:qc_diagnostic}, \ref{tab:elbow_diagnostic}, and \ref{tab:umap_diagnostic} show decision records (\textbf{Rule 1}) for the QC, dimensionality reduction, and annotation steps. \ref{stab:sup_equivalence} maps each of the ten rules to the part of these files that demonstrates it, and \ref{stab:parity} records the step-by-step parity with the official tutorial. We provide examples to apply our recommendations on information-dense visual representations such as heatmaps (\ref{sfig:heatmap} and matching \ref{stab:heatmap}) and multi-panel feature plots (\ref{sfig:featureplots} and \ref{stab:featureplots}). For low-vision readers, the figures use large fonts, a colourblind-safe high-contrast palette, and axes labelled with units. Scatter and UMAP panels are square, heatmaps are taller than wide with a scaled-expression legend, and cell-type names sit in white boxed labels at each cluster centroid rather than as small text over the points. Each figure's units also appear in its companion table.

The captions of Table~\ref{tab:elbow_diagnostic} and Table~\ref{tab:umap_diagnostic} hold the decision records for the number of principal components and for cell-type annotation, with the same fields as the quality-control record in \textbf{Rule 1}.

\begin{table}[htbp]
\centering
\caption[Non-visual companion of the PCA elbow plot and decision record for the number of principal components.]{Non-visual companion of the PCA elbow plot (Figure~\ref{fig:elbow_diagnostic}). The variance each PC explains and the running cumulative total, so the choice of components can be checked against exact numbers. Values from S1 File (\texttt{06\_pca\_variance.tsv}).
\par Decision record, with five of the required fields of~\ref{stab:adr}.
\par\textit{Question:} How many PCs should be used for clustering and UMAP?
\par\textit{Decision:} The first ten.
\par\textit{Rationale:} Variance per PC drops from 2.6\% at PC 6 to 1.7\% at PC 7 and changes little after that. The official tutorial places the elbow around PC 9 to 10 and uses ten PCs~\cite{seuratpbmc3k}.
\par\textit{Uncertainty:} PCs 7 to 12 each explain 1.3 to 1.7\% of the variance, so there is no single obvious cutoff. The tutorial reports that results often change little with 10, 15 or 50 PCs, and recommends choosing more PCs over fewer. How to choose dimensionality is outside our scope.
\par\textit{Downstream effect:} PCs 1 to 10 are used for the neighbour graph, for clustering at resolution 0.5, and for UMAP.}
\label{tab:elbow_diagnostic}
\small
\begin{tabular}{r r r}
\arthline
\textbf{PC} & \textbf{Variance \%} & \textbf{Cumulative \%} \\
\arthline
1 & 20.5 & 20.5 \\
2 & 8.2 & 28.7 \\
3 & 6.1 & 34.8 \\
4 & 5.7 & 40.5 \\
5 & 4.1 & 44.6 \\
6 & 2.6 & 47.2 \\
7 & 1.7 & 48.9 \\
8 & 1.5 & 50.5 \\
9 & 1.4 & 51.9 \\
10 & 1.4 & 53.2 \\
11 & 1.3 & 54.6 \\
12 & 1.3 & 55.9 \\
\arthline
\end{tabular}

\medskip
{\footnotesize The first ten PCs, selected for downstream analysis, account for 53.2\% of the variance among the computed PCs. Cumulative values are computed before rounding, so they need not equal the running sum of the rounded per-PC column.}
\end{table}

\begin{table}[htbp]
\centering
\caption[Non-visual companion of the cell-type UMAP and decision record for cell-type annotation.]{Non-visual companion of the cell-type UMAP (Figure~\ref{fig:umap_diagnostic}). Each cluster's size, its centroid in UMAP coordinates, and a plain-language location, so a reader knows where each population sits and how large it is without seeing the plot. Location is relative rather than absolute. A cluster falls in the right or left half, and in the upper or lower half, according to whether its centroid sits above or below the median of the nine cluster centroids on that axis. Values from S1 File (\texttt{09\_umap\_cluster\_fingerprint.tsv} and \texttt{11\_final\_celltype\_summary.tsv}).
\par Decision record, with five of the required fields of~\ref{stab:adr}.
\par\textit{Question:} Which cell type does each cluster represent?
\par\textit{Decision:} Assign the nine labels of the official tutorial to clusters 0 to 8.
\par\textit{Rationale:} Each cluster expresses its canonical markers, for example MS4A1 in 85\% of the cells of cluster 3 (B cells) and PPBP in all 13 cells of cluster 8 (platelets) (\ref{stab:heatmap} and \ref{stab:featureplots}).
\par\textit{Uncertainty:} Naive and memory CD4 T cells (clusters 0 and 2) both express IL7R and are the most connected pair of clusters in the shared nearest-neighbour graph. The naive marker CCR7 is detected in 45\% of the cells of cluster 0. Dendritic cells (32 cells) and platelets (13 cells) form small clusters.
\par\textit{Downstream effect:} The labelled object is used for the final UMAP (Figure~\ref{fig:umap_diagnostic}) and this table.}
\label{tab:umap_diagnostic}
\small
\begin{tabular}{l r c l}
\arthline
\textbf{Cell type} & \textbf{Cells} & \textbf{Centroid} & \textbf{Location} \\
\arthline
Naive CD4 T & 684 & (0.4, 5.5) & right half, upper half \\
CD14+ Mono & 481 & (-8.7, -4.7) & left half, lower half \\
Memory CD4 T & 476 & (3.3, 6.1) & right half, upper half \\
B & 344 & (3.4, -12.0) & right half, lower half \\
CD8 T & 291 & (4.4, 2.1) & right half, upper half \\
FCGR3A+ Mono & 162 & (-4.8, -4.9) & left half, lower half \\
NK & 155 & (6.5, -0.7) & right half, upper half \\
DC & 32 & (-7.8, -1.4) & left half, lower half \\
Platelet & 13 & (-10.1, 1.7) & left half, upper half \\
\arthline
\end{tabular}
\end{table}

\clearpage

\section*{Acknowledgements}

We thank Ben Carrillo for help in setting up operating systems. We also want to thank Prof. Dr. Mark Robinson for his feedback on the manuscript as well as for providing compute.

The authors used Claude Opus 5 (Anthropic) between July and August 2026 during manuscript preparation, including grammar checks, LaTeX formatting to validate against PDF/UA-2, and designing a screen-reader-friendly Figure~1, and during the preparation of the worked example, for syntax checking and output validation of the workflow. During the revision in November 2026, the authors used Claude Opus 5.5 (Anthropic) for LaTeX editing, checking new references against bibliographic records, and generating the tracked-changes versions of the manuscript. Suggested updates were reviewed by JK and IM before inclusion, and all factual claims, citations, and numerical values were checked against the primary sources and against the analysis code \cite{kientsch2026workflow}. The ten rules, the worked example, and all interpretations and conclusions are the authors' own. No AI tool contributed to study design or data generation.

\section*{Funding}

JK was supported by the Swiss National Science Foundation (grant 310030\_204648 to SN).

\section*{Competing interests}

The authors have declared that no competing interests exist.

\section*{Author contributions}

JK and IM conceived the project. JK and IM developed the software and ran the analysis. JK and IM wrote the manuscript. IM supervised the project. SN acquired funding. All authors read and approved the manuscript.

\section*{Data and code availability}

The \texttt{Snakemake} workflow and the \texttt{Quarto} source are available at \url{https://github.com/imallona/accessible_single_cell_workflow}, with a tagged release archived at Zenodo~\cite{kientsch2026workflow}, \url{https://zenodo.org/records/21535760}. The rendered reports are provided as S1 File, also archived at Zenodo~\cite{kientsch2026s1}, \url{https://doi.org/10.5281/zenodo.22015207}, a single archive holding the tagged PDF, the HTML page, and the result tables as TSV files. The example uses the 10x Genomics PBMC 3k Seurat vignette~\cite{tenxpbmc3k,seuratpbmc3k}.

\bibliographystyle{unsrt}

\clearpage

\section*{Supporting information}

\noindent\textbf{S1 File.} Accessible single-cell analysis of the 10x Genomics PBMC 3k dataset, also archived at Zenodo~\cite{kientsch2026s1}. A \texttt{Seurat} workflow, driven by \texttt{Snakemake} and rendered with \texttt{Quarto}, in which every figure is paired with a table of its values, a numeric summary, and the underlying data as a downloadable file. The archive holds the result tables as TSV files and the report rendered in two formats: a tagged PDF/UA-1 file and an HTML page. We provide both formats because they serve different needs: the tagged PDF is a portable, self-contained document that meets the PDF/UA standard, while the HTML page links directly to each result table and is generally easier to navigate by heading and by keyboard. \ref{stab:sup_equivalence} maps each rule to the part of this file that applies it.

\noindent\textbf{S1 Appendix.} Supplementary tables and figures, included after the reference list: the fields of a non-visual analytical decision record (Table~S1), the rule applications to the example workflow (Table~S2), commonly used screen readers (Table~S3), the object-state history, analytical parameters and parity with the \texttt{Seurat} vignette for the worked example (Tables~S4 to S6), excerpts of the worked example's output directory, accessibility checks and pinned environment (Figures~S1 to S3), and the marker heatmap and feature plots with their non-visual companions (Figures~S4 and S5, Tables~S7 and S8).

\clearpage
\section*{S1 Appendix. Supplementary tables and figures}

\section*{Supplementary tables}
\SupTable{stab:adr} Fields of a non-visual analytical decision record. The definitions apply to any analysis; the example column gives a value for every field of the quality-control decision of the worked example.

\begin{longtable}{P{0.20\textwidth} P{0.14\textwidth} P{0.27\textwidth} P{0.27\textwidth}}
\arthline
\textbf{Field} & \textbf{Status} & \textbf{Definition} & \textbf{Example: quality control in S1 File} \\
\arthline
\endfirsthead

\multicolumn{4}{l}{\textit{Continued from previous page.}} \\
\arthline
\textbf{Field} & \textbf{Status} & \textbf{Definition} & \textbf{Example: quality control in S1 File} \\
\arthline
\endhead

\arthline
\multicolumn{4}{r}{\textit{Continued on next page.}} \\
\endfoot

\arthline
\endlastfoot

\texttt{record\fnbrk id}
& Required
& A unique identifier for the decision record, stable across reruns.
& \texttt{qc\_01} \\

\texttt{workflow\fnbrk step}
& Required
& The analysis step the decision belongs to.
& Quality control \\

\texttt{diagnostic\fnbrk question}
& Required
& The explicit question the checkpoint answers.
& Which cells should be kept for downstream analysis? \\

\texttt{visual\fnbrk checkpoint}
& Required
& The conventional visual object that would usually support the decision.
& QC violin plot (Figure~\ref{fig:qc_diagnostic}) \\

\texttt{input\fnbrk object}
& Required
& The object or dataset inspected, with its assay or layer where relevant and its state at the time.
& \texttt{Seurat} object after creation; RNA assay, counts layer; 2{,}700 cells \\

\texttt{source\fnbrk table}
& Required
& The structured, inspectable file from which the diagnostic was generated.
& \texttt{02\_qc\fnbrk distribution\fnbrk before\fnbrk filtering.tsv} \\

\texttt{quantitative\fnbrk evidence}
& Required
& The numerical evidence supporting the decision.
& Median 816 genes (IQR 690 to 952); median 2.0\% mitochondrial; maxima 3{,}400 genes and 22.6\% mitochondrial \\

\texttt{decision}
& Required
& The decision taken as a direct result of the checkpoint.
& Keep cells with more than 200 and fewer than 2{,}500 genes and less than 5\% mitochondrial counts \\

\texttt{rationale}
& Required
& Why the decision follows from the evidence.
& Removes the high-gene and high-mitochondrial tails; parity with the official tutorial \\

\texttt{uncertainty}
& Required
& Any ambiguity or unresolved concern.
& Fixed thresholds are heuristics; the lower gene bound removes no cells \\

\texttt{alternatives\fnbrk considered}
& Recommended
& Other plausible choices that were evaluated.
& Outlier detection at three median absolute deviations: low tails of log genes and log UMIs, high tail of mitochondrial percentage \\

\texttt{sensitivity\fnbrk check}
& Recommended
& Whether the decision is stable under reasonable parameter changes.
& The alternative, computed for this table from \texttt{02\_qc\fnbrk per\fnbrk cell\fnbrk before\fnbrk filtering.tsv}, flags 173 cells, including 57 of the 62 removed, but also 10 of the 13 platelets \\

\texttt{downstream\fnbrk effect}
& Required
& Later steps that use the result of the decision.
& 2{,}638 cells passed to normalisation \\

\texttt{accessibility\fnbrk output}
& Required
& The non-visual outputs for the checkpoint.
& Table~\ref{tab:qc_diagnostic}; alt text of Figure~\ref{fig:qc_diagnostic}; object fingerprint (\ref{stab:objstate}) \\

\texttt{code\fnbrk location}
& Required
& The script, notebook or code chunk that generated the diagnostic, so another analyst can rerun it.
& Chunks \texttt{02-qc-accessible-\allowbreak summary-before} and \texttt{02-apply-\allowbreak qc-filter} in \texttt{pbmc3k\_acc.qmd} \\

\texttt{software\fnbrk environment}
& Recommended
& Relevant software, package and annotation versions; this matters most when decisions depend on tool-specific output.
& R 4.5.3, \texttt{Seurat} 5.5.0 (\texttt{envs/single\_cell.yaml}) \\

\texttt{review\fnbrk status}
& Recommended
& Whether the decision has been checked, and by whom. Suggested values: \texttt{draft}, \texttt{checked}, \texttt{revised}, \texttt{final}.
& \texttt{revised} \\

\texttt{reviewer\fnbrk notes}
& Optional
& Free-text comments from a collaborator or reviewer; useful when a decision remains plausible but uncertain.
& Peer review asked why the thresholds are appropriate and how uncertain the decision is \\

\end{longtable}
\clearpage
\SupTable{stab:sup_equivalence} Rule applications to the example workflow (S1 File).
\begin{longtable}{P{0.15\textwidth} P{0.45\textwidth} P{0.32\textwidth}}
\arthline
\textbf{Rule} & \textbf{Update} & \textbf{Evidence} \\
\arthline
\endfirsthead

\arthline
\endlastfoot

1. Treat plots as decision records, not illustrations
& Each figure is followed by a table of its values, a one-line numerical conclusion, and a link to the data; a figure index lists them all with their alternative text.
& \texttt{12\_figure\fnbrk accessibility\fnbrk index.tsv}; per-figure alt text in the report \\

2. Take AI descriptions with a grain of salt
& Alternative text is written from each plot's data and code, and the report shows that code-centric description.
& \texttt{fig-alt} fields in the \texttt{Quarto} source \\

3. Set up an accessible environment first
& The whole workflow runs from a shell, and each code chunk can be sent line by line from an editor to the terminal.
& \texttt{Snakefile}; interactive fallback in the \texttt{Quarto} source \\

4. Lean on literate programming for reproducibility
& One \texttt{Quarto} document holds prose, code, and output in small chunks; a \texttt{Snakemake} workflow runs it and a \texttt{conda} environment pins every version.
& \texttt{pbmc3k\_acc.qmd}; \texttt{Snakefile}; \texttt{envs/single\_cell.yaml} \\

5. Structure data and metadata for non-visual exploration
& Every result is a tab-separated file with a named identifier column and a lowercase, numbered filename, all listed in an output index.
& results tables; \texttt{00\_output\_index.tsv} \\

6. Facilitate data inspection and fingerprinting
& A one-line object fingerprint at each checkpoint replaces \texttt{str()}, quality control is reported as medians, interquartile ranges, and thresholds, and long output is wrapped rather than hidden in a scroll box.
& \ref{stab:objstate}; \texttt{custom.scss}; colour stripping in \texttt{Snakefile}; README accessibility notes \\

7. Publish in accessible formats
& The report opens with screen-reader navigation notes and a glossary, and renders as a tagged PDF/UA-1 (validated by \texttt{veraPDF}) and an HTML page (checked by \texttt{axe} against WCAG).
& \ref{sfig:accessible_rendering}; S1 File; \texttt{custom.scss}; \texttt{axe.yml} \\

8. Collaborate with accessibility in mind
& The analysis is encoded in a version-controlled code repository, and output filenames start with a pointer to the generating code so exchanged files stay traceable.
& the repository \cite{kientsch2026workflow}; file-naming convention; \texttt{Snakefile} logs; \texttt{README.md} \\

9. Build community and shared infrastructure
& The repository is a small, MIT-licensed, reusable template for an accessible vignette.
& repository \cite{kientsch2026workflow} and \texttt{LICENSE} \\

10. Incorporate accessibility to make FAIR research fair
& The same plain-text outputs built for non-visual use also serve every reader: the tables and logs are diffable and archivable, the scripted figures regenerate after a data update, and the numeric summaries can be audited without rerunning the analysis.
& whole workflow; plain-text results and logs \\

\end{longtable}

\clearpage

\SupTable{stab:screenreaders} Commonly used screen readers and their features.
\medskip

{\small\singlespacing\noindent
\begin{tabularx}{\textwidth}{l l l X X l X}
\arthline
\textbf{Screen reader} & \textbf{OS} & \textbf{Availability} & \textbf{Coverage} & \textbf{Cost} & \textbf{Language} & \textbf{Other} \\
\arthline
\texttt{VoiceOver} & macOS & Native & Broad within the Apple ecosystem & No additional cost & Multilingual & Nonstandard key bindings \\
\texttt{Narrator} & Windows & Native & Partial but improving & No additional cost & Multilingual & Driver needed for braille \\
\texttt{Orca} & Linux & Native & Partial for GUIs & Free, open source & Multilingual & Community driven and scriptable \\
\texttt{NVDA} & Windows & Dedicated & Broad & Free, open source & Multilingual & Portable via USB without installation \\
\texttt{JAWS} & Windows & Dedicated & Broadest & Commercial & Multilingual & Feature rich \\
\arthline
\end{tabularx}
}

\clearpage

\SupTable{stab:objstate} Object-state history. A compact fingerprint of the \texttt{Seurat} object at each checkpoint (from \texttt{12\_object\_state\_history.tsv}), transposed so the four checkpoints run across and the state fields run down.
\medskip

{\small
\begin{tabular}{P{0.15\textwidth} P{0.14\textwidth} P{0.18\textwidth} P{0.18\textwidth} P{0.19\textwidth}}
\arthline
\textbf{Field} & \textbf{After creation} & \textbf{After PCA} & \textbf{After clustering} & \textbf{After annotation} \\
\arthline
cells & 2700 & 2638 & 2638 & 2638 \\
genes & 13714 & 13714 & 13714 & 13714 \\
active assay & RNA & RNA & RNA & RNA \\
assay layers & counts & counts, data, scale.data & counts, data, scale.data & counts, data, scale.data \\
variable features & 0 & 2000 & 2000 & 2000 \\
reductions & none & pca & pca, umap & pca, umap \\
graphs & none & none & RNA\_nn, RNA\_snn & RNA\_nn, RNA\_snn \\
identity levels & 1 & 1 & 9 & 9 \\
active identities & pbmc3k & pbmc3k & 0--8 & Naive CD4 T, CD14+ Mono, Memory CD4 T, B, CD8 T, FCGR3A+ Mono, NK, DC, Platelet \\
\arthline
\end{tabular}
}

\clearpage
\SupTable{stab:params} Parameters. The analytical parameters used, so the run can be reproduced; the recorded software environment is in \texttt{12\_session\_info.txt}. Imported from \texttt{00\_parameters\_used.tsv}.
\begin{filecontents*}{data/00_parameters_used.tsv}
parameter	value
input directory	data/filtered_gene_bc_matrices/hg19
output directory	results/pbmc3k_acc
Seurat version	5.5.0
minimum cells per gene	3
minimum features per cell at object creation	200
QC minimum features	> 200
QC maximum features	< 2500
QC maximum mitochondrial percentage	< 5
normalization method	LogNormalize
normalization scale factor	10000
variable features	2000
PCA dimensions used downstream	1:10
clustering resolution	0.5
figure DPI	300
\end{filecontents*}
\csvreader[separator=tab, respect all,
  longtable={P{0.40\textwidth} P{0.50\textwidth}},
  table head={\arthline \textbf{Parameter} & \textbf{Value} \\ \arthline \endhead},
  late after line={\\}, late after last line={\\ \arthline}]{data/00_parameters_used.tsv}{}{\csvcoli & \expandafter\seqsplit\expandafter{\csvcolii}}

\clearpage

\SupTable{stab:parity} Parity with the Seurat vignette. Every analytical stage and displayed plot of the official \texttt{Seurat} PBMC 3k tutorial and its non-visual counterpart. Stages whose plot is also shown in this article name the corresponding figure. The same content is in \texttt{12\_official\_core\_parity.tsv} (S1 File).
\begin{longtable}{P{0.40\textwidth} P{0.50\textwidth}}
\arthline
\textbf{Stage or output} & \textbf{Accessibility addition} \\
\arthline
\endhead
Read10X and CreateSeuratObject & Object-state and sanity-check tables \\
Sparse count-matrix example & Long-format count table \\
Dense-versus-sparse memory comparison & Explicit memory warning \\
Mitochondrial percentage and metadata example & Captioned metadata and QC summaries \\
QC violin plot & Alternative text and numerical distribution table (Figure~\ref{fig:qc_diagnostic}) \\
Two QC scatter plots & Alternative text and threshold table \\
Official QC filter & Per-cell pass/fail flags and before-after summary \\
LogNormalize & Layer-state explanation \\
Two-panel variable-feature plot & Complete ranked variable-feature table \\
Scale all genes & Object-state checkpoint \\
RunPCA and print top loadings & Embeddings, loadings, and top-loading tables \\
VizDimLoadings & Alternative text and loading summary \\
PCA DimPlot & Alternative text and complete coordinates \\
PC1 DimHeatmap & Alternative text and loading table \\
PC1 to PC15 DimHeatmap & Alternative text and loading table \\
ElbowPlot & Numerical variance table (Figure~\ref{fig:elbow_diagnostic}) \\
FindNeighbors and FindClusters & Cluster-size and assignment tables \\
RunUMAP and cluster DimPlot & UMAP coordinates plus SNN fingerprint and narrative \\
FindMarkers cluster 2 & Downloadable complete result table \\
FindMarkers cluster 5 versus clusters 0 and 3 & Downloadable complete result table \\
FindAllMarkers positive markers & Complete and filtered marker tables \\
ROC marker example & Downloadable ROC table \\
Normalised MS4A1 and CD79A VlnPlot & Per-cluster mean, median, and detection table \\
Raw-count NKG7 and PF4 VlnPlot & Raw-count mean, median, and detection table \\
Nine-gene FeaturePlot & Highest-expression cluster table (\ref{sfig:featureplots}) \\
Top-marker DoHeatmap & Exact heatmap gene table and narrative summary (\ref{sfig:heatmap}) \\
Official canonical cluster map & Numerical canonical-marker evidence \\
Labelled final UMAP & Cell-count and marker-evidence table (Figure~\ref{fig:umap_diagnostic}) \\
Styled UMAP export object & PNG, PDF, and JPEG copies \\
Final RDS and session information & Output index, parameters, and accessible session log \\
\arthline
\end{longtable}

\clearpage
\section*{Supplementary figures and companion tables}
\singlespacing

\begin{figure}[H]
\small
\begin{Verbatim}[frame=single,fontsize=\footnotesize]
$ ls results/pbmc3k_acc/tables/
00_output_index.tsv
00_parameters_used.tsv
02_qc_distribution_before_filtering.tsv
02_qc_failure_summary.tsv
04_variable_features_ranked.tsv
06_pca_variance.tsv
08_cluster_sizes.tsv
09_umap_coordinates.tsv
10_all_positive_cluster_markers.tsv
11_final_celltype_summary.tsv
12_figure_accessibility_index.tsv
12_object_state_history.tsv
... (35 tables in total)
\end{Verbatim}
\SupFig{sfig:tables_listing} Part of the worked example's output directory. Each diagnostic is a named, plain-text table a screen reader can open and read row by row. The numeric prefix orders the files by workflow step, and the most distinguishing part of the name comes first, so the directory listing is itself navigable. The full directory holds 35 tables (S1 File).
\end{figure}

\clearpage
\begin{figure}[H]
\small
\begin{Verbatim}[frame=single,fontsize=\footnotesize]
// custom.scss: link colour, 8.5:1 contrast on white (above the 7:1 WCAG AAA threshold)
$link-color: #0b4a9c;

# axe.yml: run the axe engine on the served HTML to flag accessibility faults
format: {html: {axe: {output: document}}}

# the qmd emits a tagged PDF/UA-1 pdf; veraPDF checks it on every build
format: {typst: {pdf-standard: ua-1}}
\end{Verbatim}
\SupFig{sfig:accessible_rendering} How the worked example makes its own outputs accessible (S1 File). A high-contrast link colour meets WCAG AAA at 8.5:1 on white, the \texttt{axe}~\cite{axecore} engine checks the served HTML for accessibility faults, and the report is emitted as a tagged PDF/UA-1 file, produced with \texttt{typst}~\cite{typst}, that \texttt{veraPDF}~\cite{verapdf} validates on every render. The full settings are in \texttt{reports/custom.scss} and \texttt{reports/axe.yml}.
\end{figure}

\clearpage
\begin{figure}[H]
\small
\begin{Verbatim}[frame=single,fontsize=\footnotesize]
# envs/single_cell.yaml: pinned software, installed into an isolated env
channels:
  - conda-forge
  - bioconda
dependencies:
  - r-base=4.5.3
  - r-seurat=5.5.0
  - bioconductor-scran=1.38.0
  - bioconductor-singlecellexperiment=1.32.0
  - quarto=1.9.38
  - r-knitr=1.51
  - r-dplyr
  - r-ggplot2
  - r-ggrepel
  - r-patchwork

# one command installs the environment and runs the whole workflow
$ snakemake --cores 1 --use-conda
\end{Verbatim}
\SupFig{sfig:reproducible_env} The pinned environment and the single command that reproduces the worked example (S1 File). \texttt{Snakemake} installs the listed versions into an isolated \texttt{conda} environment and then runs the analysis and renders the report, so the run reproduces from a clean checkout without manual setup.
\end{figure}
\clearpage
\begin{center}
\artifactgraphic[width=\linewidth,height=0.92\textheight,keepaspectratio]{A heatmap displays up to ten positive markers with average log2 fold change above 1 for each cluster.}{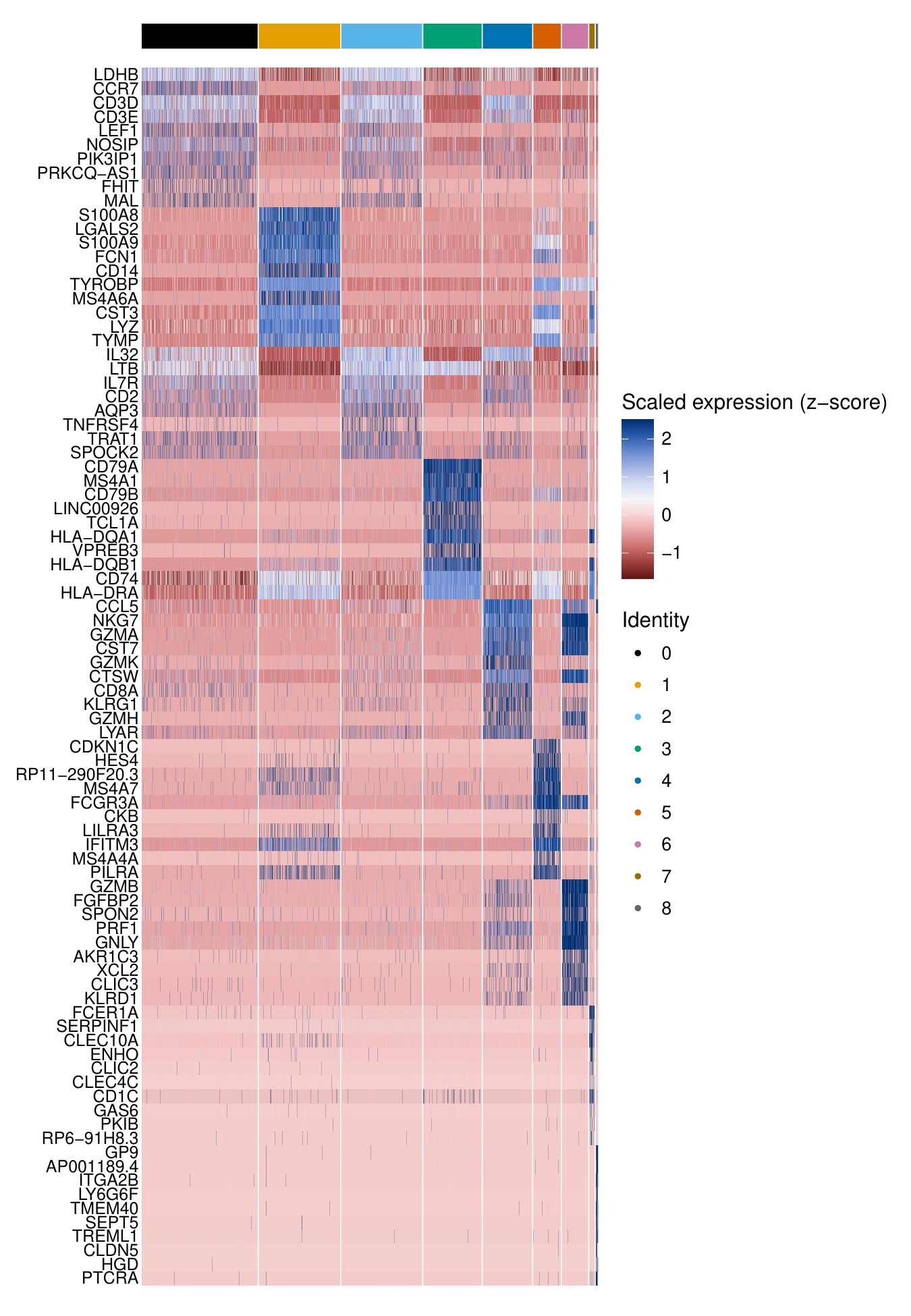}
\end{center}
\SupFig{sfig:heatmap} Marker heatmap. It shows up to ten positive markers (average log2 fold change above 1) per cluster, with scaled expression (z-score) encoded as colour. Its non-visual companion is \ref{stab:heatmap}.

\medskip

\SupTable{stab:heatmap} Non-visual companion of the marker heatmap (\ref{sfig:heatmap}). For each cluster, the top marker (lowest adjusted p-value), its log2 fold change, and the percentage of cells expressing it inside the cluster (\% in, the \texttt{pct.1} field) and in all other cells (\% out, the \texttt{pct.2} field). The full set is in \texttt{10\_top10\_heatmap\_markers\_per\_cluster.tsv} (S1 File).
\begin{center}
\small
\begin{tabular}{r l r r r}
\arthline
Cluster & Top marker & log2FC & \% in & \% out \\
\arthline
0 & LDHB & 1.2 & 91 & 59 \\
1 & S100A8 & 6.6 & 98 & 12 \\
2 & IL32 & 1.3 & 95 & 46 \\
3 & CD79A & 6.9 & 94 & 4 \\
4 & CCL5 & 3.3 & 98 & 23 \\
5 & CDKN1C & 5.4 & 51 & 1 \\
6 & GZMB & 6.0 & 96 & 7 \\
7 & FCER1A & 7.6 & 81 & 1 \\
8 & GP9 & 11.5 & 92 & 0 \\
\arthline
\end{tabular}
\end{center}

\clearpage
\begin{center}
\artifactgraphic[width=\linewidth,height=0.92\textheight,keepaspectratio]{Nine UMAP panels show the canonical PBMC markers MS4A1, GNLY, CD3E, CD14, FCER1A, FCGR3A, LYZ, PPBP, and CD8A.}{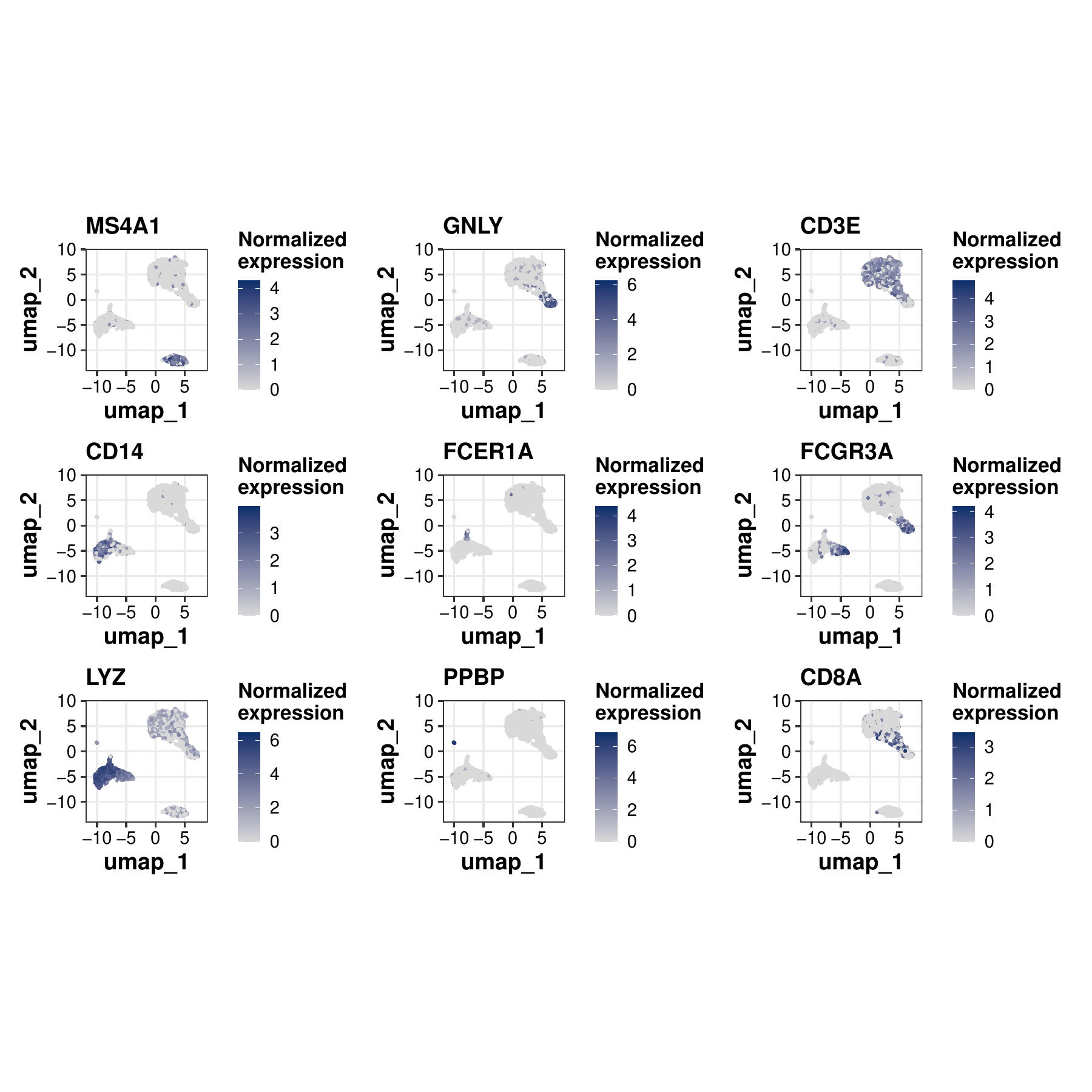}
\end{center}
\SupFig{sfig:featureplots} Marker feature plots. Nine UMAP panels show canonical marker expression, with normalised expression per cell encoded as colour intensity. Its non-visual companion is \ref{stab:featureplots}.

\medskip

\SupTable{stab:featureplots} Non-visual companion of the marker feature plots (\ref{sfig:featureplots}). For each marker, its highest-expressing cluster, that cluster's log2 fold change for the marker, and the percentage of the cluster's cells in which it is detected (\% detected, the \texttt{pct.1} field). Log2 fold changes are from \texttt{10\_all\fnbrk positive\fnbrk cluster\fnbrk markers.tsv} and expression from \texttt{10\_marker\fnbrk expression\fnbrk normalized\fnbrk by\fnbrk cluster.tsv} (S1 File).
\begin{center}
\small
\begin{tabular}{l l r r}
\arthline
Marker & Highest in & log2FC & \% detected (\%) \\
\arthline
MS4A1 & cluster 3 & 5.7 & 85 \\
GNLY & cluster 6 & 6.0 & 96 \\
CD3E & cluster 2 & 1.0 & 83 \\
CD14 & cluster 1 & 6.0 & 67 \\
FCER1A & cluster 7 & 7.6 & 81 \\
FCGR3A & cluster 5 & 4.1 & 98 \\
LYZ & cluster 1 & 4.7 & 100 \\
PPBP & cluster 8 & 11.1 & 100 \\
CD8A & cluster 4 & 3.3 & 49 \\
\arthline
\end{tabular}
\end{center}

\end{document}